\documentstyle[prd,aps]{revtex}

\begin{document}

\draft

\title{\bf Qualitative analysis of dissipative cosmologies}
\author{A. Di Prisco\footnote{On leave}}
\address{Escuela de F\'{\i}sica, Facultad de Ciencias,\\ Universidad Central 
de Venezuela, Caracas, Venezuela}
\author{L. Herrera\footnote{e-mail: lherrera@gugu.usal.es}}
\address{Area de F\'{\i}sica Te\'orica, Facultad de Ciencias, \\
Universidad de Salamanca, 37008 Salamanca, Spain}
\author{J. Ib\'a\~nez\footnote{e-mail: wtpibmej@lg.ehu.es}}
\address{Departamento de F\'{\i}sica Te\'orica,\\ 
Universidad del Pa\'{\i}s Vasco, Apdo. 644, 48080 Bilbao, Spain}
           
\date{\today}

\maketitle

\begin{abstract}
The evolution of an homogeneous and isotropic dissipative fluid is analyzed
using dynamical systems techniques. The dissipation
is driven by bulk viscous pressure and the truncated Israel-Stewart theory
is used. Although
almost all  solutions inflate, we show that only few of them can be
considered as physical
solutions since the dominant energy condition is not satisfied.

\end{abstract}


\section{Introduction}

There has been a renewed interest in the study of dissipative cosmologies
related to
the existence of inflationary solutions. As a matter of fact, dissipative
processes are thought to be present in any
realistic theory of the evolution of the universe.
The simplest model of dissipative fluid is that which
assumes the existence of bulk viscosity only. At a phenomenological level,
bulk viscosity
may be associated with particle production \cite{Z-H},\cite{barrow}. Also,
qualitatively, bulk viscosity
may be understood as the macroscopic expression of microscopic frictional
effects that appear in mixtures \cite{I-Z}.
 The easiest way to include bulk viscosity
effects
is through
the Eckart's theory \cite{eck}, which assumes that the bulk viscous
pressure is proportional
to
the expansion. Several authors have used that theory to investigate the
effects of viscosity
on  cosmological models \cite{viscos}.

As it is well known, however, Eckart's theory
suffers
important drawbacks \cite{h-s} and a more complete theory  must be used.
One of the proposed
theories is the Israel-Stewart theory \cite{is}   which complies with
causality and
stability. The non-linear terms of that theory are often neglected giving
place to the so
called
truncated theory. Although the truncated theory may lead to a different
behaviour as compared
with the full theory, we shall for simplicity base our analysis on the
truncated version of the Israel-Stewart approach.
 This truncated teory  was  firstly used by Belinskii et al. \cite{b} in a
cosmological context and after that by many authors ( see \cite{Mr} and
references therein).
 The  flat FRW models were studied in ref. \cite{p} and has been
extended
recently to spatially curved solutions \cite{p1}. In these last works the
system of field
equations was reduced to a second order differential equation and the
stability of the
stationary solutions was investigated by using a Lyapunov function.
A different and more powerful approach was given by Coley and van den
Hoogen \cite{col} who
used
dynamical
system techniques to analyze the asymptotic behaviour of the governing
field equations. They
used dimensionless equations of state that
make the equation for the expansion to decouple from the system, obtaining
a two dimensional system of differential equations.

Since these models were originally
studied,
it was soon
realized that the type of equations of state used in them was  determinant
in verifying whether
the
models
underwent inflation. It is therefore important to study the effect that
other reasonable
equations of
state will produce in the qualitative behavior of the solutions.

In this paper we analyze the asymptotic behaviour of  isotropic and spatially homogeneous
models with a fluid with bulk viscosity. It has been pointed out \cite{h-s}
that although bulk
viscosity is
almost vanishing, in both the ultrarelativistic and the Newtonian regimes,
its dynamical effects can
not
be
neglected. On the other hand the evolution equations given by this type of
matter can be used to
model another
kind of sources (e.g. the string dominated universe described by Turok
\cite{tur} or even a
scalar
field) The
equations of
state we assume are those introduced by Belinskii {\it et al.} \cite{b}.
Unlike the case  studied in ref. \cite{col}, now the expansion does not
decouple and we should  deal with the
complete three dimensional system, except for a particular value of the
parameters. It is important to stress that the choice of the equations
of state could  dramatically change the behaviour of the model and, therefore,
it is worthwhile to study the implications of  different reasonable
equations of state.

\section{Field equations}

Let us consider an homogeneous and isotropic spacetime with an imperfect
fluid moving
orthogonally
to the hypersurfaces of constant curvature. The source is a fluid with bulk
viscosity:
\begin{equation}
T_{ab}=\rho u_a u_b+ (p+\sigma)(g_{ab}+u_a u_b),
\end{equation}
where $\sigma$ is the viscous pressure $p$ is the thermodynamic pressure
and $\rho$ is the
energy density. The Einstein field equations and the conservation equations
are written as:
\begin{eqnarray}
\dot\rho & = & -3H(\rho+p+\sigma) \\
\dot H & = & -H^2-\frac{1}{6}(\rho+3p+3\sigma) \\
3H^2 & = & \rho- \frac{3k}{a^2},\qquad (k=\pm 1, 0),
\end{eqnarray}
where $a$ is the scale factor and $H\equiv \frac{\dot a}{a}$ is the Hubble
parameter. We
assume
an evolution equation for the viscous pressure $\sigma$ given by the so
called truncated
Israel-
Stewart theory (see \cite{m} and references therein):
\begin{equation}
\sigma+\tau\dot\sigma=-3\xi H.
\end{equation}
where $\xi$ is the coefficient of bulk viscosity and $\tau$ is the
relaxation time
($\xi>0, \tau>0$). Although eq.(5)(strictly speaking) is valid only on the
assumption that the fluid is close to
equilibrium ($|\sigma|\ll p$)
we will assume, for simplicity, that it describes the evolution of the
viscous pressure even
for
regimes
far from equilibrium. A more complete description would require taking into
account the
full transport
equation of the Israel-Stewart theory. We assume a barotropic equation
of state $p=(\gamma-1)\rho,\;
0<\gamma\le 2$ and
the following relations:
\begin{equation}
\xi=\alpha \rho^m,\qquad \tau=\frac{\xi}{\rho},\;\;\; m, \alpha>0
\end{equation}
which were  used by Belinskii  et al. \cite{b}.

In order to study the system of equations we define
adimensional
variables and a new
time:
\begin{equation}
\Omega\equiv\frac{\rho}{3H^2}, \qquad \Sigma\equiv\frac{\sigma}{H^2},
\qquad H(t) dt=d\tau.
\end{equation}
Now the equations (2), (3) and (5) reduce to
\begin{eqnarray}
\Omega' & = & (\Omega-1)[\Omega(3\gamma-2)+\Sigma]\\
\Sigma' & = & -9\Omega-\Sigma\left[\frac{(3\Omega)^{1-m}}{\alpha}H^{1-2m}-
2-(3\gamma-2)\Omega-\Sigma\right]\\
H' & = & -H\left[1+\frac{3\gamma-2}{2}\Omega+\frac{1}{2}\Sigma\right],
\end{eqnarray}
where $'$ means derivative with respect to $\tau$. In addition to the above
equations we have the Friedmann equation (4):
\begin{equation}
H^2(\Omega-1)=\frac{k}{a^2}.
\end{equation}
We will assume $H$ positive in the open and flat models,  so the system
(8)-(10) is well
defined. This is not true in the closed case, for which $H$ could become
zero. So in this case
the system  should be understood as describing the evolution of closed
models at early
times when the
expansion is positive.

Before studying the qualitative behavior of the system it is important to
consider
the energy conditions. The strong energy condition (\textsc{sec}) implies
that $\rho+3p_{eff}\ge 0$,
where $p_{eff}=p+\sigma$. In terms of the new variables this condition is
written as
\begin{equation}
(3\gamma-2)\Omega+\Sigma\ge 0.
\end{equation}
The violation of this condition implies that the scale factor accelerates
and the solution
inflates. The dominant energy condition (\textsc{dec}) requires
$\rho+p_{eff}\ge 0$ or:
\begin{equation}
3\gamma \Omega+\Sigma\ge 0.
\end{equation}
Matter not verifying this condition is considered unphysical, futhermore,
it seems
that the violation of this condition could lead to the violation of the
generalized
second  law of thermodynamics \cite{barrow}.

\section{Qualitative behaviour}

The line $\Omega=1$ is an invariant space and divides the phase space of
the system into three invariant sets: $\Omega>1$ corresponding to closed
models, $\Omega<1$ corresponding to open models and $\Omega=1$ to
flat
FRW solutions. Besides these invariant spaces, $H=0$ is another
invariant set that splits again the phase space into another three
invariant sets. However we do not consider the $H<0$ space  for we assume
that in early times $H>0$.

When $m=1/2$ the equation (10) decouples from the two other equations.
Therefore the system becomes a two dimensional system which
can be studied much more
easily. To investigate the qualitative features of the system we
will start with this case and then with the complete system.

\subsection{$m=1/2$}

In this case the system reduces to a two dimensional system of the form:
\begin{eqnarray}
\Omega' & = & (\Omega-1)[\Omega(3\gamma-2)+\Sigma]\\
\Sigma' & = & -9\Omega-\Sigma\left[\frac{\sqrt{3\Omega}}{\alpha}-
2-(3\gamma-2)\Omega-\Sigma\right]
\end{eqnarray}
There are four equilibrium points:
\begin{eqnarray}
P^{\pm} & : & \Omega=1,\;
\Sigma=\Sigma^{\pm}=\frac{1}{2}\left[\left(\frac{\sqrt{3}}{\alpha}
-3\gamma\right)\pm\sqrt{\left(\frac{\sqrt{3}}{\alpha}
-3\gamma\right)^2+36}\,\right] \nonumber \\
P_3 & : & \Omega=0,\;\Sigma=0,  \\
P_4 & : &
\Omega=\Omega_0\equiv\left(\frac{\alpha}{\sqrt{3}}\frac{6\gamma+5}{3\gamma-
2}\right)^2,\; \Sigma=-(3\gamma-2)\Omega,\;\;\;\; \gamma > \frac{2}{3}.\nonumber
\end{eqnarray}
The stability of these points is given by the sign of the eigenvalues of
the matrix of the linearized system around each point. The point $P_3$
has two eigenvalues of different signs, so it is a saddle
point. The two eigenvalues of the point $P^{+}$ are positive and,
therefore, it is a source point. The point $P^{-}$ has its two
eigenvalues negative provided $\alpha>\alpha_0$, where
\begin{equation}
\alpha_0\equiv\frac{\sqrt{3} (3\gamma-2)}{6\gamma+5},
\end{equation}
being, therefore, an attractor. When $\alpha<\alpha_0$ this point becomes a
saddle.
The behaviour of points $P_4$ and $P^{-}$ are opposite: when $\alpha>\alpha_0$,
$P_4$ is a saddle but is an attractor when $\alpha<\alpha_0$. Let us note
that since
$\alpha>0$ the condition $\alpha<\alpha_0$ implies $\gamma>2/3$ and
$\Omega_0<1$.
Therefore in this case
the only attractor is the point $P_4$. When $\alpha=\alpha_0$ both points,
$P_4$ and
$P^{-}$ coincide.

The exact solutions corresponding to each points are the following:
$P_3$ is the Milne universe for which $a(t)=t$. At the points $P^{\pm}$ we
have
\begin{equation}
a(t)=a_0 t^{H_0},\qquad \rho(t)=\frac{3H_0^2}{t^2},\qquad
\sigma(t)=\frac{\Sigma^{\pm}
H_0^2}{t^2},\qquad H_0\equiv \frac{2}{3\gamma+\Sigma^{\pm}}.
\end{equation}
At $P^{+}$ we have $H_0<1$ so the solution is a FRW non inflating model. At
$P^{-}$ the solution
inflates when $\alpha>\alpha_0$ $(H_0>1)$ which is equivalent to say that the
solution does not verify the strong energy condition (12). The
solution at $P_4$ is:
\begin{equation}
a(t)=a_0 t,\qquad
\rho(t)=3\left(\frac{6\gamma+5}{3\gamma-2}\;\frac{\alpha}{\sqrt{3}}
\right)^2\frac{1}{t^2},\qquad \sigma(t)=-\frac{3\gamma-2}{3}\rho,
\end{equation}
and the deceleration parameter $q=0$.

It is important to check whether the dominat energy condition (13) is
satisfied by the solutions. This condition is always satisfied at the
points $P^{+}$ and $P_4$. It is easy to see that when  the point $P^{-}$ is
an attractor
($\alpha>\alpha_0$) then eq. (13) is verified as long as
$\gamma>\alpha/\sqrt{3}$.
When $P^{-}$ is a saddle, $\alpha<\alpha_0$ means that $\gamma>\alpha/\sqrt{3}$
and the \textsc{dec} is verified at this point. In Table I the main results
concerning points
$P_4$ and $P^{-}$ are summarized.

To conclude with the description of these solutions we will compactify the
phase space
in order to analyse the behaviour at the infinity. To do this we introduce
polar
coordinates: $\Omega=\overline{r} \cos\theta$ and $\Sigma=\overline{r}
\sin\theta$.
To make $\overline{r}$ finite we define
\begin{equation}
r=\frac{\overline{r}}{1+\overline{r}},
\end{equation}
and a new time $\overline{\tau}$:
\begin{equation}
\frac{d\tau}{d\overline{\tau}}=1-r.
\end{equation}
With these definitions the system (18)-(19) can be written as
\begin{eqnarray}
\frac{dr}{d\overline\tau} & = & (1-r)\left\{r^2[(3\gamma-
2)\cos\theta+\sin\theta] -\frac{r}{\alpha}\sqrt{3r(1-
r)\cos\theta}\sin^2\theta\right.\nonumber \\
 & & \left.+(1-r)[2r-3\gamma r\cos^2\theta-
10r\cos\theta\sin\theta]\right\} \nonumber\\
\frac{d\theta}{d\overline\tau} & = & (1-r)[-
9\cos^2\theta+\sin^2\theta+3\gamma\cos\theta\sin\theta]\nonumber\\
 & & -\frac{1}{\alpha}\sqrt{3r(1-
r)\cos\theta}\sin\theta\cos\theta.
\end{eqnarray}
We deduce from the above equations that the circle $r=1$ consists of
equilibrium points.
Close to the circle we get
\begin{equation}
\frac{dr}{d\overline\tau}\sim (1-r)r^2[(3\gamma-
2)\cos\theta+\sin\theta].
\end{equation}
When $3\gamma\Omega+\Sigma>0$ the above derivative is positive and,
then, the points in the circle are attractors, while when
$3\gamma\Omega+\Sigma<0$ the derivative is negative and the
points are sources.

In Fig.1 we have plotted integral curves of the system (22) for different
values
of the parameters. The dashed curve represents the invariant space
$\Omega=1$. Points to the left
of
each vertical lines violate respectivaly each of the energy conditions.
Curves that start or end at $\theta=\pm \pi/2$ are unphysical since
$\Omega$ becomes negative and that means that at some time of their
evolution the
energy density is negative.

In Fig.1(a) the orbits starting at $P^{+}$ and going to the point
$P^{-}$, which is the only attractor for these values of the parameters,
enter into
the region where the strong energy condition is violated, so the solutions
start inflating.
Since $P^{-}$ violates the \textsc{dec}, the solutions enter into
a region where this condition is violated being, therefore, unphysical. So,
we see that in
this
case the only physical solutions are those that start at $P^+$ and go to
infinity. These
physical
solutions do not inflate. In Fig.1(b) we have plotted orbits corresponding
to values of the
parameters such that the point $P_4$ is a saddle point. The main features
in this case are
similar to that showed in Fig.1(a). The main difference is that now, since
the attractor
$P^-$ verifies the \textsc{dec}, then there are physical solutions ending
at this point. Finally in
Fig.1(c) we have plotted orbits corresponding to the case for which the
attractor is the point
$P_4$. What is remarkable in this case is that there are solutions that
after undergoing
inflation
their final state is a universe with $\Omega<1$ showing that the presence
of a state of
inflation
in the universe does not imply necessarily that $\Omega=1$ \cite{ellis}.

Let us notice from these figures that if we assume a solution starting
close to the
equilibrium
regime,  i.e. $|\sigma|\ll\rho$ or $|\Sigma|\sim 0$ then, at the end of the
evolution,
$|\Sigma|$
is of the same order than $\Omega$ which is out of the range of validity of the theory.
This  is the so called "runaway" solution \cite{h-s}
. There are
however solutions that  fulfil this requirement. If we take
$\alpha\ll 1$ then $\Sigma^-\sim -3\sqrt{3}\alpha+O(\alpha^2)$ and, in this
case, we can choose the parameters in such
a way that
$P^-$ is an attractor that verifies \textsc{dec} and the physical pressure
$p+\sigma$ of the solution is positive.
So, solutions starting with $|\Sigma|\sim 0$
evolve towards $P^-$,
being close to  equilibrium

\subsection{$m\neq 1/2$}

In this case the evolution is described by the three-dimensional system
(8)-(10).
Although it is more difficult to visualize the evolution, the behaviour is
similar to that
described
in the previous section. To further simplify the equations we define a new
variable
\begin{equation}
h=H^{1-2m}.
\end{equation}
The system (8)-(10) is cast into a simpler form
\begin{eqnarray}
\Omega' & = & (\Omega-1)[\Omega(3\gamma-2)+\Sigma]\\
\Sigma' & = & -9\Omega-\Sigma\left[\frac{(3\Omega)^{1-m}}{\alpha}h-
2-(3\gamma-2)\Omega-\Sigma\right]\\
h' & = & -(1-2m)h\left[1+\frac{3\gamma-2}{2}\Omega+\frac{1}{2}\Sigma\right],
\end{eqnarray}

The equilibrium points of the above system are:
\begin{eqnarray}
P^{\pm} & : & \Omega=1,\;h=0,\;
\Sigma^{\pm}=\frac{3}{2}\left(-\gamma\pm\sqrt{\gamma^2+4}\,
\right)
\nonumber \\
P_3 & : & \Omega=0,\; h=0, \;\Sigma=0,  \\
P_4 & : & \Omega=1\;h=\frac{3^m \alpha}{\gamma},\;
\Sigma=-3\gamma.\nonumber
\end{eqnarray}
In addition to these points there is another one, $P_5$, which is valid
only when $m>1/2$:
\begin{equation}
P_5\; :\; \Omega=1,\; H=0 \,(h=\infty),\; \Sigma=0.
\end{equation}
Let us note that when $m>1/2$ the equilibrium points, except $P_4$ and
$P_5$, have
$H=\infty$.

As before, by looking at the eigenvalues of the linearized system we obtain the
stability of these points. The points $P^{\pm}$ have eigenvalues
\begin{equation}
3\gamma-2+\Sigma^{\pm},\qquad 3\gamma+2\Sigma^{\pm},\qquad
-\frac{1-2m}{2}(3\gamma +
\Sigma^{\pm}).
\end{equation}
For $P^{+}$ the two first eigenvalues are positive and the third is
negative when $m<1/2$
and positive when $m>1/2$. Thus, this point is a saddle point if $m<1/2$
and a source if
$m>1/2$. For $P^-$ the two first eigenvalues are negative and the third is
positive when
$m<1/2$ and negative when $m>1/2$. Thus, this point is a saddle point if
$m<1/2$ but is an
attractor if $m>1/2$.

The eigenvalues of the point $P_3$ are:
\begin{equation}
\frac{1}{2}(4-3\gamma+3\sqrt{\gamma+4}),\qquad
\frac{1}{2}(4-3\gamma-3\sqrt{\gamma+4}),\qquad -(1-2m).
\end{equation}
The first eigenvalue is positive and the second negative. The third is either
negative or positive depending on the value of $m$. Thus this point is a saddle
point. Finally  $P_4$ eigenvalues are
\begin{equation}
-2,\qquad \frac{3}{2\gamma}[-(\gamma^2+1)\pm\sqrt{(\gamma^2+1)^2-2(1-2m)}\;].
\end{equation}
If $m<1/2$ the three above eigenvalues are negative and, then, $P_4$ is an
attractor. But if $m>1/2$ one of the eigenvalues is positive and $P_4$
turns to be a
saddle point.

To analize the stability of the point $P_5$ we change to spherical
coordinates: $\Omega=
\overline r\sin\theta
\cos\phi,\;\Sigma=\overline r\sin\theta\sin\phi, h=\overline r\cos\theta$
and we compactify
the phase  space by using the coordinate $r$ given by (20) and the time
$\overline\tau$ defined by (21). Close to the point $P_5$, ($r\sim 1,\;
\sin\theta\sim 1-r,\; \phi=0$) we obtain:
\begin{equation}
\frac{dr}{d\overline\tau}\sim -\frac{1}{2}(1-2m)3\gamma (1-r)^2.
\end{equation}
The above expression is positive when $m>1/2$ and therefore this point is an
attractor.

The solution described by the point $P_4$ is given by
\begin{equation}
a(t)=a_0e^{H_0t},\qquad \rho=3H_0^2,\qquad \sigma=-3\gamma H_0^2, \qquad
H_0=\left(\frac{3^m\alpha
}{\gamma }\right)^{1/(1-2m)}.
\end{equation}
This de Sitter solution was already obtained by Barrow \cite{barrow}. Since
the rest of equilibrium
points have $H=0$ (or infinity depending on the value of $m$) it is not
possible to get the
corresponding exact solutions (except for the point $P_3$ that corresponds
to the Milne solution)
We can, however, obtain approximate solutions for some of the points. When
$m>1/2$ the point $P^+$
is a source. By linearizing the system (25)-(27) around this point we get
\begin{equation}
\tau\rightarrow -\infty, \qquad H\sim H_0 e^{-\lambda \tau},\qquad
\lambda=\frac{1}{2} (3\gamma+\Sigma^+),
\end{equation}
where $H_0$ is a constant of integration. Integrating the other two
variables and using (11) to
change from $\tau$ to $t$ we obtain:
\begin{equation}
t\rightarrow 0,\qquad \rho\sim \frac{3}{(\lambda t)^2},\qquad \sigma\sim
\frac{\Sigma^+}{(\lambda
t)^2},\qquad a(t)\sim t^{1/\lambda}.
\end{equation}

To see the behavior close to the point $P_5$ we can integrate eq.(27) and
after a straightforward
calculation we obtain
\begin{equation}
t\rightarrow \infty,\qquad H\sim \frac{2}{3\gamma t}.
\end{equation}
This behaviour in the vicinity of the point $P_5$ was already obtained by
Belinskii et al. \cite{b}
and by Barrow \cite{barrow}.

As to the energy conditions, it is easy to see that both are satisfied by
the points $P^+$ and
$P_3$ and are violated by the point $P^-$. The \textsc{dec} is satisfied by
$P_4$ and $P_5$ but \textsc{sec} is
not satisfied by $P_4$ and is satisfied by $P_5$ when $\gamma\ge 2/3$.

The behavior described above is summarized in Table II.

The compactification of the phase space in the case $m\neq 1/2$ is much
more difficult than in the
$m = 1/2$ case,
 making almost impossible to get a complete description of
the system. However we can obtain the main features concerning the
qualitative evolution of
solutions from the information on the equilibrium points and the numerical
integration of the system.

In Fig.2 we have plotted the projections of the orbits, obtained by
numerical integration of
the system  (25)-(27), onto the plane ($\Omega, \Sigma$) for different
values of the parameters.
The straight lines indicate whether the solutions fulfil each of the energy
conditions. Points
below each line do not satisfy the corresponding energy condition. As in
the former case there
are solutions that enter to or come from the $\Omega<0$ region. All these
solutions are
unphysical.

Fig.2(a) corresponds to the case where the only attractor is the point $P_4$.
Besides the solutions that end at infinity, there are solutions approaching
point $P^+$, which is a saddle, then
entering into an inflationary regime and as approaching the point $P^-$,
which is a saddle as well, they become
unphysical since the dominat energy condition is violated. The values of
the parameter in
Fig.2(b) are such that there are two attractors: points $P^-$ and $P_5$.
Since in this case
$P^-$ does not satisfy the \textsc{dec}, the solutions going to this point
become unphysical. However,
solutions ending at $P_5$ could be physical. These last solutions could be
of physical interest
for they are the only solutions that could fulfil the requirement that the
viscous pressure be
much less than the equilibrium pressure during their evolution. Starting at
points for which
$\Sigma\sim 0$ they evolve towards solutions with $\Sigma=0$.

\section{Conclusions}

In this paper we have analyzed the qualitative behaviour of a particular
type of dissipative fluids
in FRW universes using the truncated Israel-Stewart theory. There have been
a previous \cite{p}
analysis of the same system and with the same equations of state but only
for the zero curvature
FRW universes that was thereafter extended to the general case \cite{p1}. In those analysis the
system of equations was reduced to a second order differential equation
whose stationary solutions
were studied by using a Liapunov function. That method does not
discriminate, however, between
different equilibrium points. We have seen in this paper that when $m\neq
1/2$ almost all the
equilibrium points have $H=0$ and all of them were reduced to only one
stationary solution in the
mentioned papers.

An important feature that emerges from our analysis is that for most of the
equilibrium points 
either the dominat energy condition or the positivity of the energy density
are not
satisfied. Solutions evolving towards these equilibrium points,
therefore, should be ruled out as unphysical. That means that in
order to obtain physical solutions we have to be very careful with both the
initial conditions and the
values of the parameters.  With respect to inflation, we see that,
generically,
the presence of bulk viscosity makes the solutions enter into an
inflationary regime.
Only few solutions do not inflate.

If we consider the fluid interpretation of our system of equations, we have
seen that almost all
the solutions evolve in such a way that they go outside  the regime in
which the theory is valid
(i.e. when $|\sigma | \ll p$). The only exceptions are those solutions
which evolve towards
the point $P_5$ when $m>1/2$ which describes
an inflating universe when $\gamma<2/3$ or solutions evolving towards $P^-$
with $m=1/2$ and
$\alpha$ very small.

As mentioned before we have used the truncated equation (5),
instead of the more general transport equation derived from
the Israel-Stewart theory.
As it has been shown \cite{m}, this latter equation may be written as
\begin{equation}
\tau_{*} \dot\sigma + \sigma = - 3 \xi_{*} H
\left[1+\frac{1}{\gamma c_b^2 }\left(\frac{\sigma}{\rho}\right)^2\right]
\label{A}
\end{equation}
with
\begin{equation}
c_b^2 = \frac{\xi}{(\rho + p) \tau}
\end{equation}
and
\begin{equation}
\tau_* = \frac{\tau}{1 + 3 \gamma \tau H} \; ; \;
\xi_* = \frac{\xi}{1 + 3 \gamma \tau H}.
\end{equation}
Therefore, close to equilibrium ($\sigma << \rho$), equation (\ref{A}) leads
to the truncated equation (5) with reduced
relaxation time ($\tau_*$) and bulk viscosity ($\xi_*$)
\begin{equation}
\tau_{*} \dot \sigma + \sigma = - 3 \xi_{*} H.
\label{B}
\end{equation}
The ammount of reduction depending on the size of $\xi$ relative to $H$.
Obviously if $\tau H << 1$, then $\tau_* \approx \tau$ and $\xi_* \approx \xi$.
However if $\tau H$ is close to $1$, the reduction could be significant.
On the other hand, qualitative changes in the results, outside the
quasi-equilibrium regime, are difficult to forecast without
considering specific models.

It is important to realize that the system of  equations studied in this paper,
not only describes the evolution of a
dissipative fluid
with bulk viscosity, but also the evolution of a different kind of matter. It
is
well
konwn, for instance, that when $m=2/3$ the above equations give the
evolution of a string
dominated universe
\cite{tur}. We shall see that this equivalence can be extended to a scalar
field as well.

Indeed, a scalar
field in both
FRW and Bianchi type solutions is equivalent to a perfect fluid with $p$
and $\rho$ defined as:
\begin{equation}
\rho=\frac{1}{2}\dot\phi^2+V(\phi),\qquad p=\frac{1}{2}\dot\phi^2-V(\phi),
\end{equation}
where $V$ is the scalar field potential. We can substitute in the above
expression $p$ by
$p+\sigma$. If at
the same time we consider a barotropic equation of state between the energy
density and the
equilibrium
pressure of the form $p=(\gamma-1)\rho$ then we obtain that the viscous
pressure is given by:
\begin{equation}
\sigma=\frac{2-\gamma}{2}\dot\phi^2-\gamma V.
\end{equation}
Taking derivatives with respect time and substituting into the Klein-Gordon
equation of the scalar
field
\begin{equation}
\ddot\phi+3H\dot\phi+\frac{dV}{d\phi}=0
\label{pul}
\end{equation}
we obtain that $\sigma$ satisfies an equation like (5) with  $\tau$ and
$\xi$ given by
\begin{equation}
\tau=\frac{1}{6H}, \qquad \xi=\frac{1}{3H}\left( \gamma
V+\frac{1}{3H}\frac{dV}{d\phi}\dot\phi\right).
\label{ul}
\end{equation}

Now, in dissipative systems there exists a peculiar state called ``critical
point'', which corresponds
to a specific value of a combination of thermodynamic variables. It has
been shown \cite{HeMa}, \cite{HeDP}
(and references therein) that a dissipative system in the critical point
immediately
(i.e. on a time scale of the order of the relaxation time) after its
departure from equilibrium,
behaves as if the effective inertial mass of any fluid element vanishes.

In the case of pure bulk viscosity the critical point is characterized by
\begin{equation}
\frac{\xi}{\tau}=2(\rho+p).
\label{cp}
\end{equation}
In this case however the critical point is forbidden by causality and
stability requirements,
demanding
\begin{equation}
\frac{\xi}{\tau}<(\rho+p)\left(1-c_s^2\right)
\label{con}
\end{equation}
where $c_s$ is the speed of sound (see \cite{HeMa}).

Nevertheless in order to see what the critical point may imply in terms of
a scalar field, let us assume (\ref{cp}).
Then, combining with (\ref{ul}), we obtain after an elementary algebra
\begin{equation}
\frac{1}{3H}\,\frac{dV}{d\phi}=\frac{\gamma \dot{\phi}}{2}
\label{ec}
\end{equation}
and feeding back into (\ref{pul})
\begin{equation}
\ddot\phi+3H\dot\phi \left(1+\frac{\gamma}{2}\right)=0.
\label{dp}
\end{equation}
A simple integration of (\ref{dp}) gives
\begin{equation}
\dot\phi \sim e^{-\int{3H(1+\gamma/2)dt}}.
\label{in}
\end{equation}

Let us consider two examples:
\begin{itemize}
\item de Sitter case, $H=$constant.

In this well known case, one gets from (\ref{pul})
\begin{equation}
\dot\phi \sim e^{-3H(1+\gamma/2)t}
\end{equation}
and therefore
\begin{equation}
\phi \sim \dot\phi.
\end{equation}
Then (\ref{ec}) yields
\begin{equation}
V \sim \phi^2.
\end{equation}
This kind of potential is one of the most favoured by particle theorists,
since it describes renormalizable particle theory \cite{Ry}.
\item $H \sim \frac{2}{3\gamma t}$

The asymptotic behaviour close to the point $P_5$ described above (see
eq.(37)), yields,
using (\ref{in})
\begin{equation}
\dot\phi \sim \phi^{(1+\gamma/2)}
\end{equation}
and therefore the corresponding potential is
\begin{equation}
V \sim \phi^{\gamma+2}.
\end{equation}
\end{itemize}
Thus, the critical point condition (\ref{cp}), leads, for an scalar field
interpretation of
the source term, to polinomial potentials (for the two examples considered).

It is evident, therefore, that the constraints imposed by the fluid
interpretation
can be droped if  we consider the energy momentum tensor to correspond to
another kind of matter
(different from the fluid with bulk viscosity).  In such a case the only
warnings to
bear in mind are those coming from the dominat energy condition requirement.

\acknowledgments

One of us (J.I.) is indebted to Prof. A.A.Coley and Dr. R.Sussman for 
stimulating discussions. This work has been  supported
by CICYT (Spain) grants PB96-0250 and PB96-1306.

\begin{figure}
\caption{Plots of the integral curves of the sytem when $m=1/2$. Vertical axis
represents the coordinate $r$ an the horizontal axis represents the $\theta$
coordinate. They range from 0 to 1 and from $-\pi/2$ to $\pi/2$ respectively.
Dashed line represents the curve $\Omega=1$. Points to the right of the
\textsc{sec} line
verify the strong energy condition and points left don't. The same for
\textsc{dec} line,
corresponding to the dominat energy condition. (a) In this case $\gamma=1/3$ and
$\alpha=1$, ($\alpha>\alpha_0$, $\gamma<\alpha/\sqrt{3}$). (b) Now $\gamma=2$ and $\alpha=1$,
($\alpha>\alpha_0$) and (c) $\gamma=2$ and $\alpha=0.2$, ($\alpha<\alpha_0$)}
       
\end{figure}

\begin{figure}
\caption{Projections of the integral curves of the system when $m\ne 1/2$.
Points below
the solid diagonal line does not verify the strong energy condition,
while those below
the dashed  diagonal line does not verify the dominat energy condition.
(a) In this case
$\gamma=1/3$, $m=1/4$ and $\alpha=1$. (b) Now $\gamma=1/3$, $m=0.7$ and
$\alpha=1$.}
\end{figure}

\begin{table}
\caption{Qualitative properties of points $P^-$ and $P_4$}
\begin{tabular}{ll|c|ccc}
 & & $P_4$ & \multicolumn{2}{c}{$P^{-}$} & \textsc{dec} \\
\hline
 & & & & $\gamma>\alpha/\sqrt{3}$ & Yes \\
$\alpha>\alpha_0$ & $\gamma<2/3$ &  & Attractor &
$\gamma<\alpha/\sqrt{3}$ & No \\
\hline
 & & & & $\gamma>\alpha/\sqrt{3}$ & Yes \\
$\alpha>\alpha_0$ & $\gamma>2/3$ & Saddle & Attractor &
$\gamma<\alpha/\sqrt{3}$ & No \\
\hline
$\alpha<\alpha_0$ & $\gamma>2/3$ & Attractor & Saddle &
 & Yes 

\end{tabular}

\end{table}

\begin{table}

\caption{Qualitative properties of points equilibrium points}
\begin{tabular}{l|c|c|c|c|c}

 &  $P^+$ & $P^{-}$ & $P_3$ & $P_4$ & $P_5$ \\
\hline
 $m<1/2$ & Saddle & Saddle  & Saddle  & Attractor &  \\
\hline
 $m>1/2$ & Source & Attractor & Saddle & Saddle & Attractor \\
\hline
\textsc{dec} & Yes & No & & Yes & Yes \\
\hline
\textsc{sec} & Yes & No & & No & if $\gamma\ge 2/3$, Yes 

\end{tabular}

\end{table}

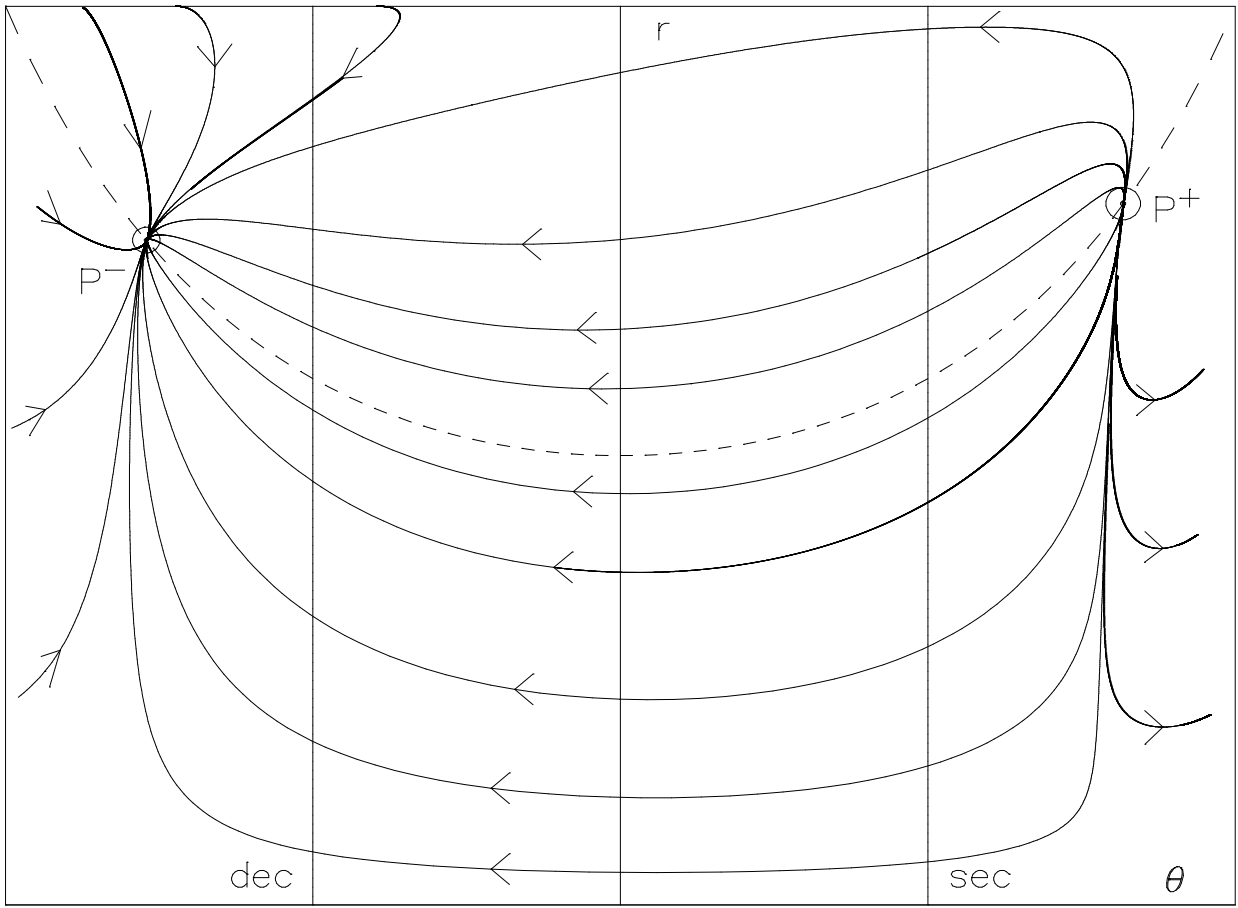
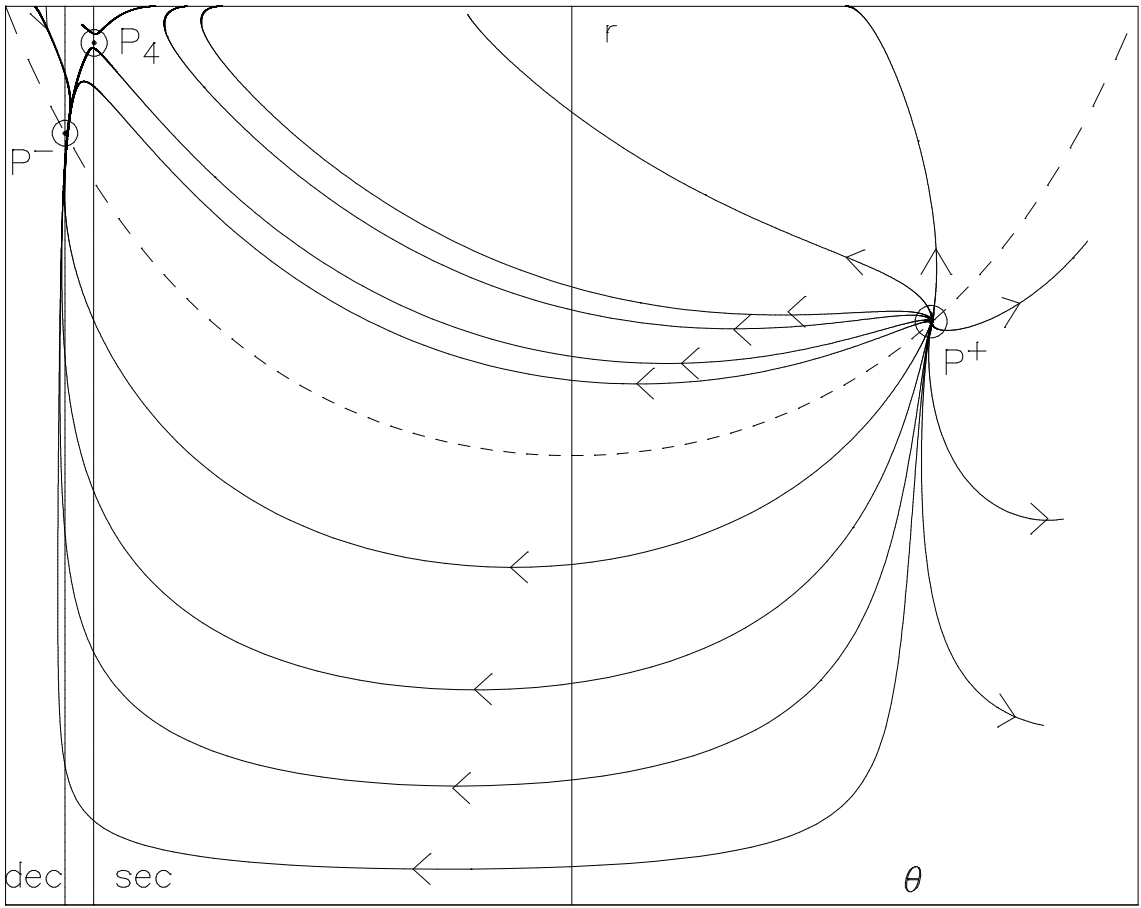
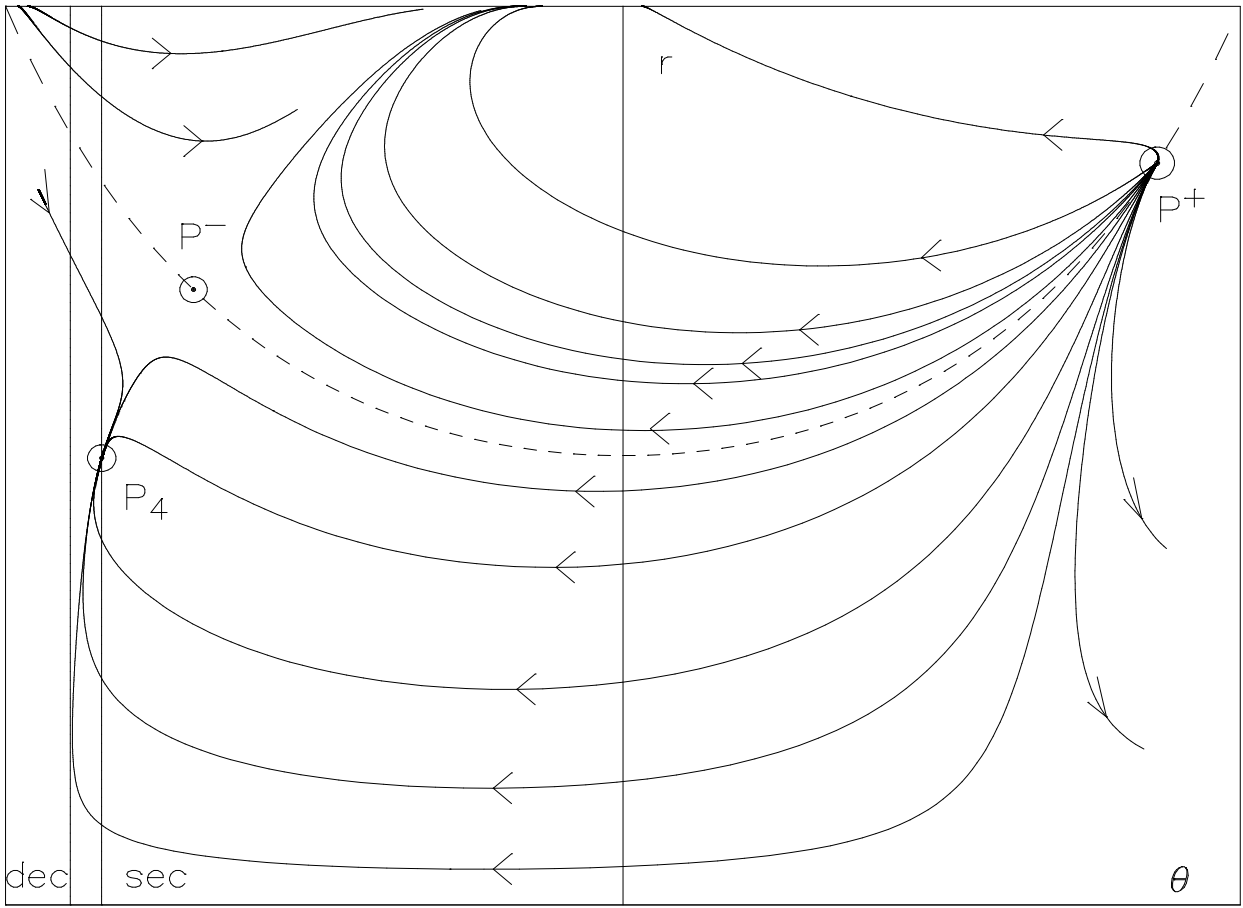
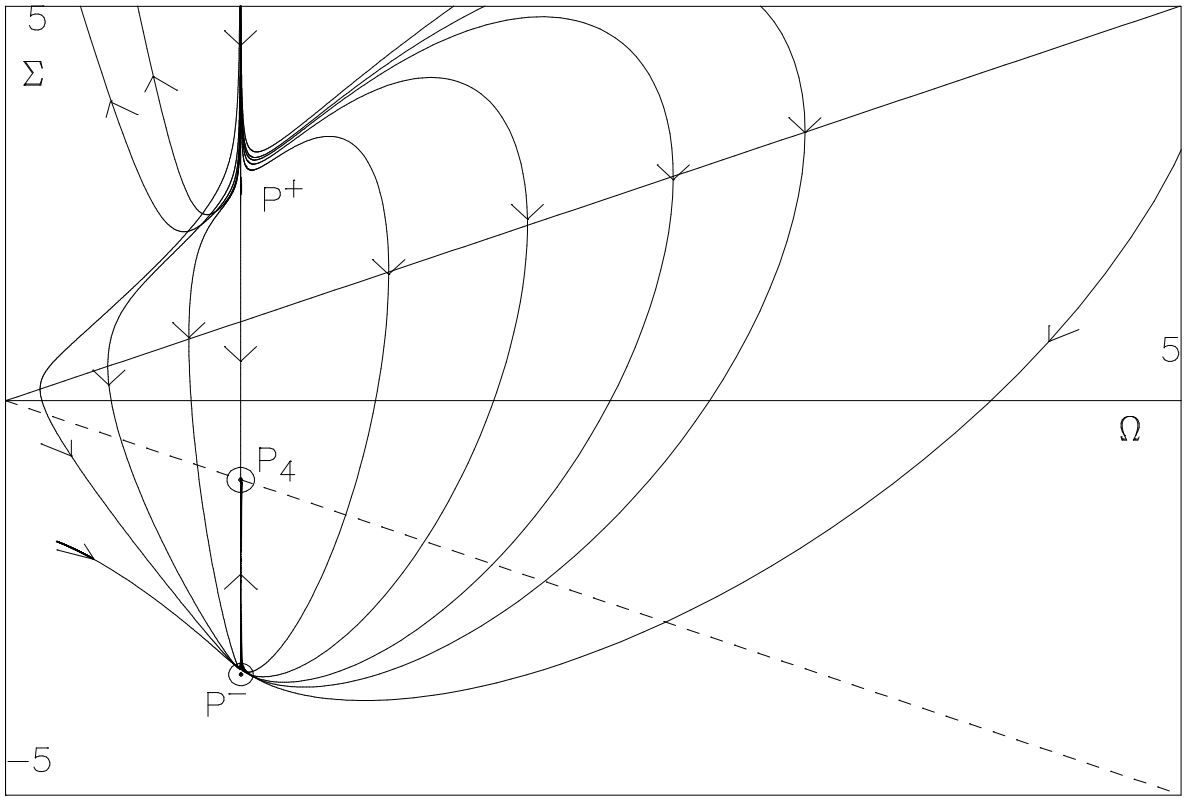
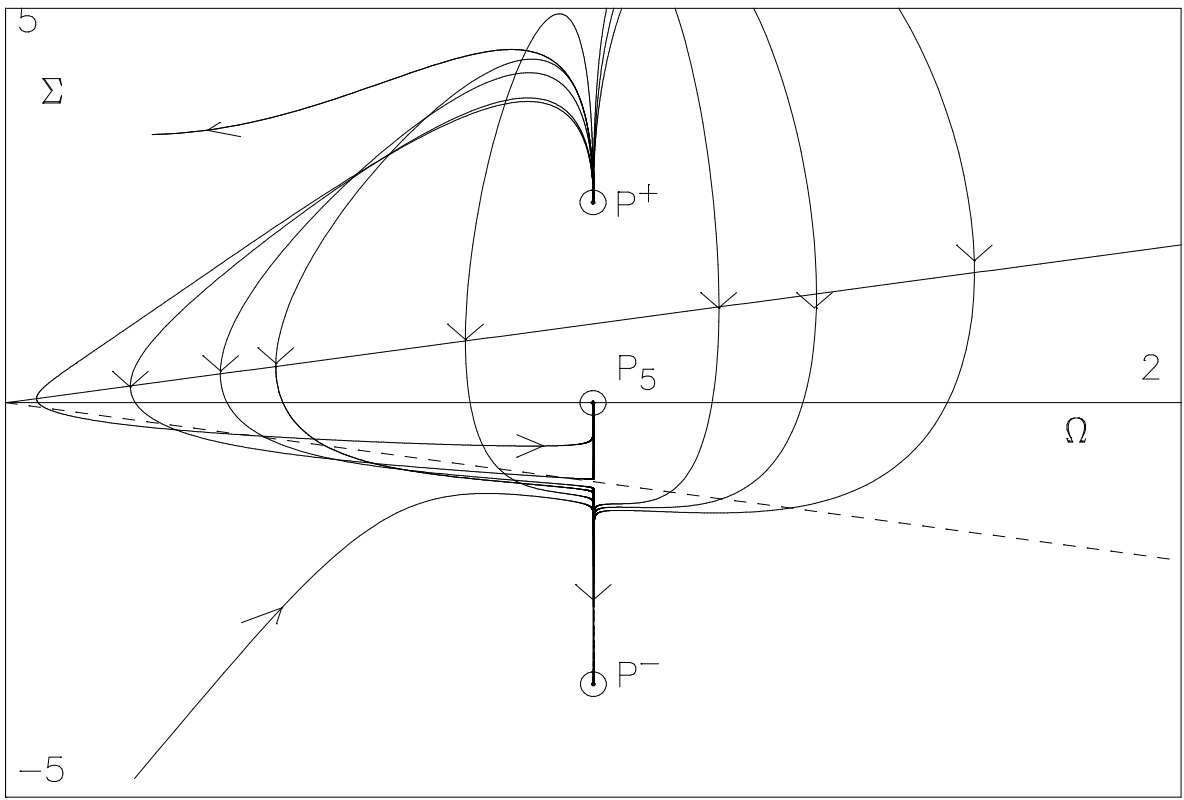
\end{document}